\documentclass[conference]{IEEEtran}
\usepackage{amsmath,amsfonts,amssymb}
\usepackage{mathtools}
\usepackage{booktabs}
\usepackage{algorithmic}
\usepackage{array}
\usepackage[caption=false]{subfig}
\usepackage{textcomp}
\usepackage{stfloats}
\usepackage{url}
\usepackage{verbatim}
\usepackage{graphicx}
\usepackage{soul}
\usepackage{xcolor}
\usepackage{balance}
\usepackage{cite}

\usepackage{tikz}

\usepackage{adjustbox}
\usepackage{bm}
\usepackage[acronym,style=alttree,nomain,shortcuts,nonumberlist,nogroupskip]{glossaries}
\newacronym{6G}{6G}{sixth generation}

\newacronym{ae}{AE}{autoencoder}
\newacronym{aoa}{AoA}{angle of arrival}
\newacronym{awgn}{AWGN}{additive white Gaussian noise}

\newacronym{ap}{AP}{access point}

\newacronym{bs}{BS}{base station}

\newacronym{ckm}{CKM}{channel knowledge map}
\newacronym{cir}{CIR}{channel impulse response}
\newacronym{csi}{CSI}{channel state information}

\newacronym{det}{DET}{detection error trade-off}

\newacronym{glrt}{GLRT}{generalized likelihood ratio test}
\newacronym{gnn}{GNN}{generative neural network}

\newacronym{ekf}{EKF}{extended Kalman filter}

\newacronym{kf}{KF}{Kalman filter}

\newacronym{isac}{ISAC}{integrated sensing and communication}

\newacronym{los}{LoS}{line-of-sight}
\newacronym{lrt}{LRT}{likelihood ratio test}
\newacronym{lt}{LT}{likelihood test}

\newacronym{ls}{LS}{least squares}

\newacronym{mimo}{MIMO}{multiple input multiple output}
\newacronym{ml}{ML}{maximum likelihood}
\newacronym{music}{MUSIC}{multiple signal classification}

\newacronym{ocsvm}{OC-SVM}{one-class support vector machines}

\newacronym{pla}{PLA}{physical layer authentication}
\newacronym{pl}{PL}{path loss}

\newacronym{rnn}{RNN}{recurrent neural network}

\newacronym{snr}{SNR}{signal to noise ratio}

\newacronym{uav}{UAV}{unmanned aerial vehicle}
\newacronym{ue}{UE}{user equipment}
\newacronym{uwb}{UWB}{ultra-wide band}

\newacronym{v2x}{V2X}{vehicle-to-everything}

\newacronym{rrc}{RRC}{root raise cosine}

\newacronym{lb}{LB}{link budget}

\newacronym{iid rv}{iid rv}{independent and identically distributed random variable}

\newacronym{ula}{ULA}{uniform linear array}  

\def\BibTeX{{\rm B\kern-.05em{\sc i\kern-.025em b}\kern-.08em
    T\kern-.1667em\lower.7ex\hbox{E}\kern-.125emX}}
\usepackage{balance}
\DeclareMathOperator*{\argmax}{arg\,max}

\usepackage{pgfplots}
\pgfplotsset{compat=1.18}
\pgfplotsset{plot coordinates/math parser=false}
\usetikzlibrary{plotmarks, arrows, decorations, spy, backgrounds, patterns, fadings, shadows.blur, shapes, positioning, calc, matrix}
\usepgfplotslibrary[colorbrewer]
\usepackage{grffile}
\pgfplotsset{plot coordinates/math parser=false}
\newlength\figurewidth
\newlength\figureheight
\begin{document}
\bstctlcite{IEEEexample:BSTcontrol}

\title{Physical Layer Authentication With Channel Knowledge Maps in Indoor Environments \vspace{0mm}}
\author{Luca Bonaventura$^\star$, Francesco Ardizzon, and Stefano Tomasin\\[2mm]
\small 
\IEEEauthorblockA{University of Padova and National Inter-University Consortium for Telecommunications (CNIT), Italy\\}
    \IEEEauthorblockA{\small $^\star$Corresponding author, email:luca.bonaventura@phd.unipd.it%
    \vspace{-6mm}}
\thanks{This work was supported by Agenzia per la cybersicurezza nazionale under the programme for promotion of XL cycle PhD research in cybersecurity - C96E24000010005. The views expressed are those of the authors and do not represent the funding institution.}}

\IEEEoverridecommandlockouts

\maketitle

\begin{abstract}
\Ac{pla} allows to authenticate the user by comparing measurements over time, assuming their time consistency or by modeling their evolution. 
However, these assumptions become problematic when devices are in motion and in indoor environments due to multipath propagation and obstructions. In this paper, we propose a \ac{pla} mechanism for moving devices in indoor environments, where multiple \acp{ap} estimate the dominant channel tap \ac{pl} and \ac{aoa} from the received signals and compare them with previously collected \acp{ckm}. Specifically, the measurements are compared to those in the neighborhood of the previously known position obtained from \acp{ckm}. A comprehensive security analysis is conducted under both random and optimal attacks. Numerical results in a representative indoor scenario, with \ac{ckm} obtained via ray tracing, validate the effectiveness of the proposed \ac{pla} approach.
\end{abstract}

\begin{IEEEkeywords}
Physical layer authentication, channel knowledge map, indoor localization.
\end{IEEEkeywords}

\glsresetall

\section{Introduction}\label{sec:intro}
\Ac{6G} networks are expected to have a dense deployment, allowing both user performance improvements and the development of new services. 
A new feature is environmental-aware communication, in which the service is improved by leveraging prior knowledge of the environment, embedding it into the communication system via \ac{ckm}~\cite{Zeng2024Tutorial}.
Motivated by the growing attention to security, in this paper, we exploit the \acp{ckm} for \ac{pla}. An overview of the \ac{pla} techniques is presented in~\cite{Zhang2024Survey}, which surveys both device fingerprinting and channel-based authentication. In this work, we focus on the latter, where the unique properties of the channel and, in particular, the propagation medium, are used to provide security. 

The key idea behind \ac{pla} verification is to determine whether a set of features extracted from channel observations matches an a priori analytical measurement model~\cite{Baracca2012Physical}. Compared with its cryptographic counterparts, \ac{pla} requires limited computational resources and does not introduce additional communication overhead; thus, it is suitable for rate-limited communication scenarios or energy-constrained devices.
However, obtaining the required a priori analytical model for verification is challenging when the user device is mobile or located in indoor environments. Indeed, in mobile scenarios, the measurement statistics vary over time, and therefore \ac{pla} the test should be adjusted accordingly. Moreover, in indoor environments, the non-\ac{los} component, which is harder to model and predict than the \ac{los} component, often becomes predominant, further complicating the model design. Signal obstructions, e.g., due to furniture or people moving in the room, further complicate this task.

Concerning environmental communications, the concept of \ac{ckm} is deliberately general, to include maps of different natures for different applications. \acp{ckm} include physical environment maps, i.e., models of the 3D environment~\cite{Esrafilian2019Learning}, as well as channel feature maps. These maps relate quantities such as \ac{aoa}~\cite{Zhao2023Common}, channel gains~\cite{Dall'Anese2011Channel, Li2022Channel}, or beams to receiver (or user) locations~\cite{Dai2024Prototyping}. 

A first example of the \ac{ckm} application includes \ac{uav} networks~\cite{Haoyun2023Channel}, where drone positions are chosen to maximize the sum of the rates between \ac{uav} and the \acrlong{bs}. 
Next, in~\cite{Dai2024Prototyping}, a beam index \ac{ckm} is used by a mmWave system to achieve training-less beamforming, considering also non-\ac{los} and dynamic conditions.

A parallel research trend investigates techniques to build the \acp{ckm}. 
In \cite{ZhangChaoyue-2025}, \ac{isac} is employed to estimate the position of a moving user in the room. The user transmits continuously while moving around the room. In turn, the receiver estimates the channel from the received signal, and combines it with the \ac{isac}-estimated user position to construct the \ac{ckm}. 
In~\cite{Zhao2023Common}, two algorithms are proposed to improve the \acp{ckm} based on \ac{aoa}. 

Finally, to the best of the authors' knowledge, only a few works explore the use of \ac{ckm} for authentication. A strategy in which the verifier performs \ac{pla} by verifying the consistency between the position reported by the user under verification and the position estimated by the verifier from the \ac{csi} is proposed in~\cite{Wang2024Knowledge}. However, the protocol assumes that the legitimate user trajectory is defined (a priori) by the verifier. 


In this work, we combine the concepts of \ac{ckm} and \ac{pla} to design a novel \ac{pla} protocol for moving devices in indoor environments. In particular, the use of \ac{ckm} eliminates the need for an a priori analytical \ac{csi} model. Specifically, we consider multiple \acp{ap} estimating \ac{pl} and \ac{aoa} from the dominant channel tap, i.e., the tap with the highest amplitude. Then, an authentication mechanism verifies, through \ac{lt} or \ac{glrt}, whether the estimates match those stored in \ac{ckm}, looking at those within the area surrounding the previously estimated location. We design different attack strategies against the proposed mechanism. These include both random and informed attacks, i.e., where the attacker exploits knowledge about the \ac{ckm} and past positions of the target user. The numerical results in a reference indoor environment confirm the effectiveness of the proposed protocol \ac{pla}.

The remainder of the paper is organized as follows. Section~\ref{sec:SystemModel} details the system model. Section~\ref{sec:overview} describes the proposed mechanism. Section~\ref{sec:results} reports the numerical results. Finally, Section~\ref{sec:conclusion} draws the conclusions.

\section{System Model}\label{sec:SystemModel}
\begin{figure}[ht!]
    \centering
    \begin{adjustbox}{width=\columnwidth, keepaspectratio}
    \input{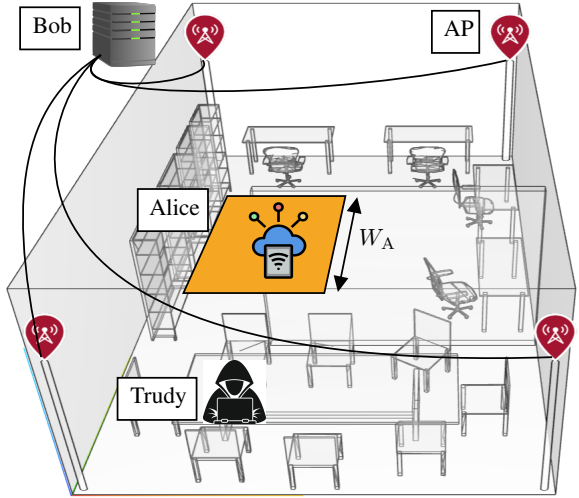}
    \end{adjustbox}%
    \caption{Schematic representation of the security scenario involving Alice, Bob, and Trudy.}
    \label{fig:generalSchemeAuth}
    \vspace{-.5cm}
\end{figure}

We consider the indoor authentication scenario shown in Fig.~\ref{fig:generalSchemeAuth}, where $N$ \acp{ap} have publicly known fixed positions and operate under centralized control, denoted as Bob. Two user devices move within the area covered by the network, namely Alice and Trudy.
We consider an authentication problem where the verifier Bob aims at deciding whether the received signal has been transmitted by Alice or by the attacker Trudy, who pretends to be Alice to intrude on the system. 

The \acp{ap} are loosely synchronized and are equipped with an \ac{ula} of $N_\mathrm{A}$ antenna elements, equally spaced by $d$. In turn, Alice and Trudy are single-antenna devices.

We assume that the communication between the \acp{ap} controlled by Bob is authenticated and integrity-protected; therefore, the attacker cannot manipulate the information shared between the \acp{ap} and Bob. Being the \acp{ap} static, this can be achieved by using a wired or fiber connection.

\subsection{Channel model}

The channel is narrow-band with the user transmitting signals at frequency $f_\mathrm{c}$ and wavelength $\lambda = c/f_\mathrm{c}$, with $c$ the speed of light. 
Alice transmits $N_{\rm p}$ publicly known pilot sequences, used by Bob for channel estimation purposes. Both the user and the \ac{ap} operate with symbol period $T_\mathrm{s}$.

Alice and Trudy move in a horizontal plane modeling the room, sampled in squares of side $W_M$. Then, each 2D position $\bm{p} = (x, y)$, represents the center of one of these squares, and is collected in $\mathcal{P}$.

We model the channel from $\bm{p}$ to the $n$-th \ac{ap} using $R_n(\bm{p})$ distinct rays, where the $i$-th ray is characterized by time of arrival, $T_{i,n}(\bm{p})$, \ac{aoa} measured in radians, $D_{i,n}(\bm{p})$, and \ac{pl}, $L_{i,n}(\bm{p})$. 
Thus, considering a transmitted power $P_{\rm T}$, the user and \ac{ap} antenna gains, $G_{\rm T}$ and $G_{\rm R}$, respectively, the power received at the \ac{ap} from the $i$-th ray is 
\begin{equation}
    P_{\mathrm{R}}\big(L_{i,n}(\bm{p})\big)= \frac{P_{\rm T}\, G_{\rm T} \, G_{\rm R}} {L_{i,n}(\bm{p}) \, L_{\mathrm{D}}}\,,
\end{equation}
where $L_{\mathrm{D}}$ accounts for all other losses not included in the \ac{pl}.

The impulse response of the broadband channel for the $a$-th element of the \ac{ula} with $a \in \{ 0, 1, \dots, N_\mathrm{A} - 1 \}$ of the \ac{ap} $n$, is 
\begin{equation}\begin{split}
\tilde{h}&_{a, n}(t,\bm{p}) = \\ &=\frac{1}{2}\sum_{i=1}^{R_n(\bm{p})} B_{a}\big(L_{i,n}(\bm{p}), \,T_{i,n}(\bm{p}), \, D_{i,n}(\bm{p}) \big)\, 
\delta(t - T_{i}) \, ,
\label{eq:cont_broadband_channel_model}
\end{split}\end{equation}
where $B_{a}(L, \,T, \, D) = \sqrt{P_{\mathrm{R}}(L)} \, \exp\left({-j 2 \pi \, f_{\rm c} \, T}\right)\,  \alpha_a(D)$, $\alpha_a(D) =\exp\left({-j2 \pi \,d \, \sin\left(D\right)\, a/\lambda}\right)$, and  $\delta(\cdot)$ is the Dirac delta function~\cite{Njeri-2014}.

Assuming that both user and \ac{ap} employ a \ac{rrc} filter for interpolation and decimation, characterized by a roll-off factor $\rho$ and a modulation interval $T_\mathrm{s}$~\cite[Ch. 7]{Benvenuto-2021}, the equivalent base-band channel is
\begin{equation}\begin{split}\label{eq:BBChannel}
    h_{a,n}(t, \bm{p}) = \frac{1}{2}\sum_{i=1}^{R_n(\bm{p})} B_{a}(L_{i,n}(\bm{p}), \,T_{i,n}(\bm{p}), \, D_{i,n}(\bm{p})) \, \\\mathrm{rcos} (t - T_{i,n}(\bm{p}), T_\mathrm{s}, \rho)\, , 
\end{split}\end{equation}
where $\mathrm{rcos}(t, T_\mathrm{s}, \rho)$ denotes the raised cosine function~\cite[Ch. 7]{Benvenuto-2021}.
We assume that the first ray, i.e., with $i=1$, is the one with the highest received power. This ray will be later used by the \acp{ap} for synchronization.

\subsection{\ac{csi} Estimation}\label{sec:measEstimation}
The proposed \ac{pla} mechanism is based on the \ac{csi} estimated by the \acp{ap}, which is obtained as follows.

Let us consider the $a$-th antenna of $n$-th \ac{ap}, performing estimation using the $m$-th received pilot signal, with $ m \in \{0,\dots, N_{\mathrm{p}}-1\}$.
Then, the dominant channel tap, i.e., the one arriving at time $T_{1,n}(\bm{p})$, $h_{a,n}\big(T_{1,n}(\bm{p}), \bm{p} \big)$, for the transmitter in position $\bm{p}$ is estimated as 
\begin{equation}\begin{split}
    \hat{g}_{a,n}^{(m)}(\bm{p}) = h_{a,n}\big(T_{1,n}(\bm{p}), \bm{p} \big) + w\, , 
\end{split}\end{equation}
where $w \sim \mathcal{CN}(0, \sigma^2)$ is the \ac{awgn}, independent at different antennas. More in detail, the estimation noise standard deviation is $\sigma = K \, \sigma_0$ where $K$ and $\sigma_0^2$ denote the noise scaling factor and the unscaled noise power, respectively. 
Hence, overall, $\hat{g}_{a,n}^{(m)}(\bm{p}) \sim \mathcal{CN}\big( h_{a,n}(T_{1,n}(\bm{p}),\bm{p}), \sigma^2 \big)\,$, 
and the \ac{ap} can then estimate the dominant channel tap, later used to infer \ac{pl} and \ac{aoa}, by collecting different estimations of $\hat{g}_{a,n}^{(m)}(\bm{p})$.
$|\hat{g}_{a,n}^{(m)}(\bm{p})|$ follows a Rician distribution 
with shape $K_{\mathrm{r},n}(\bm{p}) = h^2_{a,n}\big(T_{1,n}(\bm{p}), \bm{p}\big) / \sigma^2$, and scale parameter $\Omega_n(\bm{p})  = h^2_{a,n}\big(T_{1,n}(\bm{p}), \bm{p}\big)+\sigma^2$ that depends on \eqref{eq:BBChannel} and, in turn, the location of the user.
Then, we can estimate the \ac{pl} as~\cite{Sijbers-1998}
\begin{align}
\hat{r}_n(\bm{p}) &=\frac{P_{\rm T} \, G_{\rm T} \, G_{\rm R}}{ L_{\mathrm{D}} \,|\hat{\mu}_n(\bm{p})|^2}\, ,
\end{align}
where 
\begin{equation}
|\hat{\mu}_n(\bm{p})|^2 =  \biggr[\frac{1}{N_\mathrm{A} \, N_\mathrm{p}} \sum_{a =0}^{N_\mathrm{A}-1}\sum_{m =0}^{N_\mathrm{p}-1} |\hat{g}_{a,n}^{(m)}(\bm{p})|^2 \biggr] - \sigma^2\, .
\end{equation}

Thus, the estimated \ac{pl} at the $n$-th \ac{ap}, $\hat{r}_n(\bm{p})$, can be modeled as a random variable with mean $r_n(\bm{p})$, i.e., the true \ac{pl}, and variance $\sigma^2_{\mathrm{r},n}(\bm{p})$. In particular, by averaging over $N_{\rm A} \, N_{\rm p}$ measurements, we can leverage distribution convergence $\xrightarrow{d}$ via the central limit theorem, i.e., 
\begin{equation}
    \sqrt{N_{\rm p }N_{\rm A}}\left(\hat{r}_n(\bm{p})  - r_n(\bm{p})\right) \xrightarrow{d} \mathcal{N}\left(0, \sigma^2_{\mathrm{r},n}(\bm{p})\right)\,.
\end{equation}
Thus, $\hat{r}_n(\bm{p})\sim \mathcal{N}\left(r_n(\bm{p}), \sigma^2_{\mathrm{r},n}(\bm{p})\right)$. 

The \acrlong{ml} estimate of the \ac{aoa} can be obtained by considering an angular sampling interval $\zeta \in \{ 0, \dots, \pi\} $ and evaluating~\cite{Stoica-1989}
\begin{equation}
    \hat{\theta}_n(\bm{p}) = \argmax_\zeta \Bigg|\sum_{a=0}^{N_{\rm A}-1}\sum_{m=0}^{N_{\rm p}-1} \hat{g}_{a,n}^{(m)}(\bm{p})\, \alpha^*_a(\zeta) \Bigg| \, ,
\end{equation}
where $^*$ denotes the complex conjugate. Assuming that the angular sampling is dense, $\sigma$ is small, and $N_{\rm p}$ is large enough, the \ac{aoa} in the ${n}$-th \ac{ap} can be approximated as \footnote{For larger $\sigma$ values, a folded Gaussian distribution may be more appropriate to account for the wrapping of the angular domain.}
\begin{align}
    \sqrt{N_{\rm p}}\left( \hat{\theta}_n(\bm{p}) - \theta_n(\bm{p})\right) \xrightarrow{d} \mathcal{N}\left(0, \sigma^2_{\mathrm{\theta},n}(\bm{p})\right)\,,
\end{align}
where $\theta_n(\bm{p})$ is the \ac{aoa} of the first ray, i.e., the one associated with the highest received power.

To improve notational clarity, we denote the estimated \ac{pl} and \ac{aoa} at \ac{ap} $n$ during the authentication phase, corresponding to the position $\bm{p}_k$ at time $t_k$, as $\hat{r}^{(k)}_n$ and $\hat{\theta}^{(k)}_n$, respectively, and collected them as $\hat{{\phi}}^{(k)}_n = \{ \hat{r}^{(k)}_n , \hat{\theta}^{(k)}_n \}$. 
Finally, we concatenate all \acp{ap}’ observations collected at time $t_k$ into the vector $\hat{\bm{\phi}}^{(k)} = [\hat{{\phi}}^{(k)}_1 , \ldots, \hat{{\phi}}^{(k)}_N ]^\top$.


\subsection{Channel Knowledge Map}\label{sec:CKMDerivation}
We assume Bob to be provided with a \ac{ckm}~\cite{Zhao2023Common,Dall'Anese2011Channel, Li2022Channel} derived using any method described in Section \ref{sec:intro}, where, for each \ac{ap} $n$, a user location $\bm{p}$ is mapped to ${{\phi}}_{n} (\bm{p}) = \left\{r_{n}(\bm{p}) ,\,  \theta_{n}(\bm{p})\right\}$. The vector of \ac{ckm} at Bob is ${\bm{\phi}}(\bm{p}) = [{\phi}_{1}(\bm{p}), \dots,{\phi}_{N}(\bm{p}) ]^\top$.
The \acp{ckm} are considered to be noiseless, e.g., computed by averaging over an arbitrarily large set of measurements.

We assume Alice's starting position, i.e., at time $t_0$, to be authenticated, e.g., by using an upper-layer authentication protocol. Then, at time $t_k = k \,T_\Delta$, she moves to 
\begin{equation}
\bm{p}_k = \bm{p}_{k-1}+\bm{v}_{k-1}\, T_\Delta \, ,
\end{equation}
where $\bm{v}_k = (v_{\mathrm{x},k}, v_{\mathrm{y},k})$, with $v_{\mathrm{x},k} , v_{\mathrm{y},k} \sim\mathcal{U}[-\nu, \nu]$, $\nu$ is the maximum user velocity. 
Alice's trajectory, and thus position $\bm{p}_{k}$, is unknown to both Bob and Trudy. However, both leverage Alice's maximum velocity, $\nu$, discovering that her next position must lie within a square window of side $W_{\rm A} = \nu  \, T_\Delta$, centered at $\bm{p}_{k-1}$. The set of Alice's possible positions in such a window is denoted by $\mathcal{P}_k \subseteq \mathcal{P}$.

\subsection{Security Assumptions and Attacker Capabilities}
Attacker Trudy aims to impersonate Alice to intrude on the system.  The communication protocol's parameters, e.g., the pilot sequence, are public and used by Trudy to generate the spoofing signals. 
Additionally, we assume that Trudy is synchronized with the legitimate devices and can overshadow Alice's broadcast transmission. Thus, Bob cannot use inconsistencies between signals from different \acp{ap} for \ac{pla}.

Finally, we will analyze and design attacks considering scenarios where i) Trudy has not \acp{ckm}, and thus uses random position attacks, and ii) she has access to the maps, and therefore can select her position to increase the success probability of her attack.

\section{Proposed Authentication Mechanism}\label{sec:overview}
We now describe the proposed \ac{pla} mechanism and the Trudy's attack strategies, which will later used for the security analysis.

\subsection{Authentication Mechanism}\label{sec:AuthStrategy}

The authentication mechanism operates as follows. In a preliminary phase, Bob acquires the \acp{ckm} $\bm{\phi}(\bm{p})$, as described in Section~\ref{sec:CKMDerivation}.
Bob also knows Alice's position $p_0$ at time $t_0$. 
At time $t_{k}$, a message is received, and Bob must determine whether Alice or Trudy transmitted it. To this end, the message includes $N_{\rm p}$ pilot sequences that allow Bob to estimate the dominant tap of the channel, as described in Section~\ref{sec:measEstimation}, from which he obtains measurements $\hat{\bm{\phi}}^{(k)}$.
Now, based on the current observation $\hat{\bm{\phi}}^{(k)}$ and on the \acp{ckm} $\bm{\phi}(\bm{p})$, Bob must assess the authenticity of the received signal. A hypothesis-testing approach is adopted, later detailed in Section~\ref{subsec:Hypothesis_Testing}.

If the received signal is deemed authentic (i.e., originating from Alice), the Alice’s position is updated via a maximum a posteriori procedure as 
\begin{equation}
    \hat{\bm{p}}_{k} = \argmax_{\bm{p}\in\mathcal{P}_k}   p\left( \hat{\bm{\phi}}^{(k)}  \Big| \bm{\phi} = \bm{\phi}(\bm{p}) \right) \,,
\end{equation}
and $ \hat{\bm{p}}_{k}$ is the new position for the next authentication instant.

If the received signal is deemed not authentic, i.e., originating from Trudy, an attack is detected, the message $k$ is discarded, the estimate of Alice’s position remains unchanged.



\subsection{Hypothesis Testing}
\label{subsec:Hypothesis_Testing}
We frame the authentication problem using binary hypothesis testing, distinguishing between the legitimate  $\mathcal{H}_0$ and alternative hypotheses $\mathcal{H}_1$, i.e., when Alice or Trudy is the transmitter, respectively. 
The optimal test is the \ac{lrt}, which compares the likelihoods for $\mathcal{H}_0$ versus $\mathcal{H}_1$~\cite{Zhang2024Survey}. Still, it requires knowledge about the observation model under both legitimate and alternative hypotheses. Therefore, such a test cannot be used in a security context, where no information is provided about the attacker to the verifier. The defender must then resort to either a \ac{lt} or a \ac{glrt}~\cite{Zhang2024Survey}. The former uses the likelihood of being in a specific position in $\mathcal{P}_k$ as test function, i.e.,
\begin{equation} \begin{split} \label{eq:likelihood}
\gamma_{\mathrm{ LT}, k} &= \max_{\bm{p}\in\mathcal{P}_k}  p\left( \hat{\bm{\phi}}^{(k)}  \Big| \bm{\phi} = \bm{\phi}(\bm{p}) \right) =
\\ &= \max_{\bm{p}\in\mathcal{P}_k} \prod_{n=1}^N p\left(\hat{r}_n^{(k)} | r_n={r}_n(\bm{p})\right)   p\left(\hat{{\theta}}^{(k)}_n | {\theta}_n  = {\theta}_n(\bm{p})\right),
\end{split} \end{equation}
where we factored the terms by leveraging the independence between the \ac{pl} and \ac{aoa} measurements at different \acp{ap}.

When using the \ac{glrt}, the defender protects himself against the worst-case attack, i.e., the position more likely to be chosen by Trudy, and the test function becomes \begin{equation}\label{eq:GLRT} \begin{split}
\gamma_{\mathrm{GLRT}, k} &= \frac{\max_{\bm{p}\in\mathcal{P}_k}  p\left( \hat{\bm{\phi}}^{(k)}  \Big| \bm{\phi} = \bm{\phi}(\bm{p}) \right) }
{\max_{\bm{p} \notin \mathcal{P}_k} p\left( \hat{\bm{\phi}}^{(k)}  \Big| \bm{\phi} = \bm{\phi}(\bm{p}) \right)}  \,.
\end{split} \end{equation}
Notice that the \ac{lt} can be seen as a \ac{glrt} with an undefined alternative, where the maximum is taken over an arbitrarily large set of parameters~\cite{Ardizzon2025Relation}. Still, each approach has its advantages and disadvantages. In particular, \ac{glrt} patches the weakness of the \ac{lt} by defending against the worst-case attack. On the other hand, this makes the test more vulnerable to all attacks that are different from the worst-case attack. These considerations will be confirmed by simulations in Section~\ref{sec:results}.

In both cases, given that the test function $\gamma_k$, i.e., \eqref{eq:likelihood} or \eqref{eq:GLRT}, the authenticity is verified by deciding between
\begin{equation}\label{eq:detector}
    \hat{\mathcal{H}}_k = \begin{cases}
        \mathcal{H}_0 & \mbox{if}\quad \gamma_k\geq \xi\,,\\
        \mathcal{H}_1 & \mbox{if}\quad \gamma_k < \xi\,,
    \end{cases}
\end{equation}
where $\xi$ is a user-defined threshold for labeling the signal as legitimate or false. 
The performance of the test is evaluated considering the false alarm, i.e., the probability of labeling as fake the legitimate signal, $P_{\rm FA} = P[\hat{\mathcal{H}_1}|\mathcal{H}_0]$ and the missed detection, i.e., the probability of labeling as legitimate a signal transmitted by Trudy, $P_{\rm MD} = P[\hat{\mathcal{H}_0}|\mathcal{H}_1]$.


In detail, the likelihoods in \eqref{eq:likelihood} and \eqref{eq:GLRT} are computed as follows. As discussed in Section~\ref{sec:measEstimation}, both \ac{pl} and \ac{aoa} are Gaussian distributed, thus considering for instance, the \ac{pl},  
\begin{equation}\begin{split}\label{eq:PLlikelihood}
  & p\Big(\hat{r}_n^{(k)} | r_n={r}_n(\bm{p})\Big)   =\\
    &=\begin{cases}
        P_{\emptyset, n}, \quad \,\, &\mbox{if } \hat{r}_n^{(k)} = {r}_n(\bm{p})=\emptyset \,,\\  
        0, \quad \quad &\mbox{if }\hat{r}_n^{(k)} \neq r_n(\bm{p}) = \emptyset\,\,,  \\
        0, \quad \quad &\mbox{if } \hat{r}_n^{(k)} \neq \hat{r}_n = \emptyset\,,  \\
                (1 - P_{\emptyset, n}) f (\hat{r}_n^{(k)};{r}_n(\bm{p}),\hspace{-.3cm}& \sigma_{\mathrm{r},n}(\bm{p}) ),\mbox{otherwise} ,
    \end{cases}
\end{split}\end{equation}
where $f(x; \mu ,\sigma)$ is the pdf of the Gaussian distribution with mean $\mu$ and standard deviation $\sigma$, and  $ P_{\emptyset,n}$ counts the fraction of how many positions in $\mathcal{P}_k$ have signal obstructions for \ac{ap} $n$.
An expression analogous to \eqref{eq:PLlikelihood} is computed for $p\left(\hat{{\theta}}_n | {\theta}_n(\bm{p})\right)$.

\subsection{Attack Strategies}\label{sec:attackStrategy}
We consider three attack strategies: random position, informed, and full map attacks.
\paragraph*{Random Position Attack} Trudy is in a random position, $\bm{p}_\mathrm{T}$, inside the room but not close to Alice, i.e., $\bm{p}_\mathrm{T} \in \mathcal{P}$ but $\bm{p}_\mathrm{T} \notin \mathcal{P}_k$, and broadcasts pilot signals to the \acp{ap}.
\paragraph*{Informed Attack} Trudy, initially placed in position $\bm{p}_\mathrm{T}$, knows Alice's position, the whole \ac{ckm}, and takes advantage of the check \eqref{eq:likelihood} to adjust her position. However, we assume that she can pick only positions close to the current location, i.e., the positions in the set $\mathcal{P}_\mathrm{T} \subset \mathcal{P}$ collecting the positions in the square of size $W_\mathrm{T} \times W_\mathrm{T}$ centered at $\bm{p}_\mathrm{T}$.
Thus, she moves to the best position in $\mathcal{P}_\mathrm{T}$, calculated as  
\begin{equation}\label{eq:informedAtt}
    \bm{p}^\star_{\mathrm{T},k} = \argmax_{\bm{p} \in \mathcal{P}_\mathrm{T}} p\left( \bm{\phi}(\bm{p})\Big| \bm{\phi} = \bm{\phi}(\bm{p}) \right)  \,.
\end{equation}

\paragraph*{Full Map Attack} Trudy is capable of moving instantly and in advance to any position in the room, not in $\mathcal{P}_k$. Thus, she selects the position from which the attack is more likely to succeed as 
\begin{equation}
\label{eq:optimalAtt}
    \bm{p}_{\mathrm{T},k} = \bm{p}^\star_k = \argmax_{\bm{p} \notin \mathcal{P}_k} p\left( \bm{\phi}(\bm{p})  \Big| \bm{\phi} = \bm{\phi}(\bm{p}) \right)  \,.
\end{equation}
Notice that this represents an upper bound to attack \eqref{eq:informedAtt}, where $\mathcal{P}_\mathrm{T}$ is extended to the entire indoor environment. In turn, the random position attack is a trivial informed attack where $W_\mathrm{T} = 1$. 
\section{Numerical Results}\label{sec:results}

The \ac{ckm} are generated using the MATLAB ray-tracing tool with the channel model of Section~\ref{sec:measEstimation}.
The ray-tracer considers a room with dimensions of $8 \times 5 \times 3\,\text{m}^3$, with the furniture shown in Fig.~\ref{fig:generalSchemeAuth}. A similar environment has been used in~\cite{Hamza-2022}.

Next, we run a Monte Carlo simulation in which, for each set of the considered parameters, we 
\begin{enumerate}
    \item set $\bm{p}_{k-1}$ as a random (sampled) position in the room. 
    \item choose $\bm{p}_{k}$ as a random position\footnote{We include only feasible positions, e.g., outside walls.} within $\mathcal{P}_k$, i.e., square window of side $W_{\rm A}$ centered at $\bm{p}_k$,
    \item set $\bm{p}_\mathrm{T}$ using one of the attacks described in Section \ref{sec:attackStrategy},

\end{enumerate}
We remark that, in this work, we have simulated the authentication of a single step, i.e., from $t_{k-1}$ to $t_k$, rather than the whole trajectory, leaving the study, for instance, of the authentication performance at subsequent steps after the attack, as a future work.

Table~\ref{tab:simParameters} collects the simulation parameters, chosen to be compliant with IEEE 802.11 for noise scaling factor $K = 10^{-3}$, considering an extra margin in the link budget $L_{\rm D}$ to account for all the receiver non-idealities, e.g., device losses and array-antenna element misalignment.
\begin{table}
    \centering
    \caption{Default Simulation Parameters}
    \label{tab:simParameters}
    \begin{tabular}{ccc} \toprule
        \textbf{Parameter} & \textbf{Description} & \textbf{Value}  \\ \midrule
        $N$ & Number of \acp{ap}& $4$\\
        $N_{\rm A}$ & Number of Rx antennas & $4$\\
        $N_{\rm p}$ & Number of pilots symbols & $10$\\
        $(P_{\mathrm T})_{\rm dBm}$ & User Tx power & $15\,$dBm\\
        $(G_{\mathrm{T}})_{\rm dBi}$ & User antenna gain & $2\,$dBi\\
        $(G_{\mathrm{R}})_{\rm dBi}$ & \ac{ap} single antenna element & $2\,$dBi\\
        $(L_{\mathrm{D}})_{\rm dB}$ & Other attenuation losses& $5\,$dB\\
        $(\sigma^2_{{0}})_{\rm dBm}$ & Unscaled noise power \ac{ap}& $-35\,$dBm\\
        $f_{\rm c}$ & Carrier frequency& $2\,$GHz\\ 
        $d$ & Antenna elements spacing & $\lambda/2$\\
        $T_{\rm s}$ & Symbol period & $100\,$ns\\
        $\rho$ & \ac{rrc} roll-off factor&$0.9$\\
        $\zeta_\mathrm{s}$ & Angular sampling interval& $0.1\,$$^\circ$ \\
        $W_{\rm M}$ & \ac{ckm} squares size & $0.1\,$m\\
        $W_{\rm A}$ & Alice movement window& 
        $7 \, W_M$\\ 
        $W_{\rm T}$ & Trudy movement window& $7 \, W_M$\\ \bottomrule
    \end{tabular}
    \vspace{-.5cm}
\end{table}

\subsection{Authentication Results}
\begin{figure*}
    \centering
    \subfloat[$K = 0.1$, varying $N$. \label{fig:singleAuth}]{\input{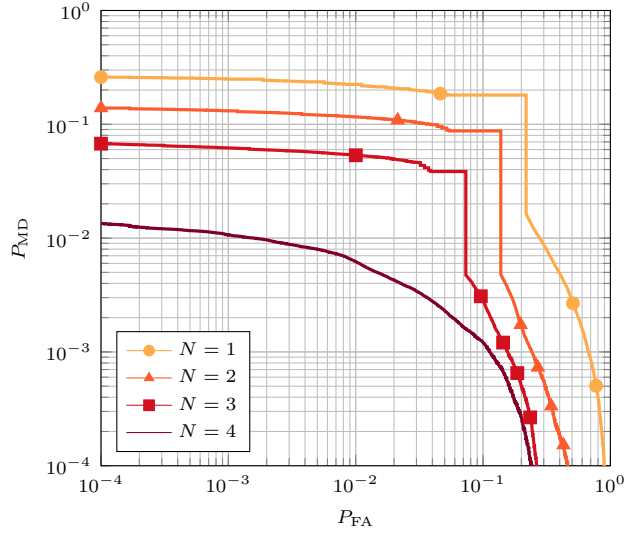}}
    \subfloat[Varying $K$, fixed $N=4$.\label{fig:resMulti}]{\input{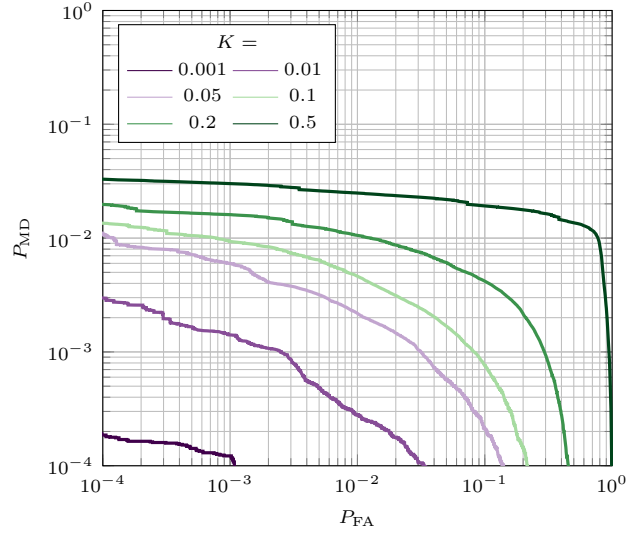}}  
    \caption{\Ac{det} curves against the random position attack for various estimation noises values, $\sigma = K \sigma_0$, and available \acp{ap}, ${N}$.}
    \label{fig:authGuessing}
    \vspace{-.5cm}
\end{figure*}

First, we investigate the effects of the number of available \acp{ap} (Fig.~\ref{fig:singleAuth}) and noise (Fig.~\ref{fig:resMulti}) on the authentication performance of the \ac{lt} against the random position attack, by considering the \ac{det} curves, i.e., the missed detection as a function of the false alarm probability. 
As expected, the detection improves as the estimation noise decreases and $N$ increases. However, increasing $N$ brings more benefits to authentication performance. Indeed, having fewer \acp{ap} means having more obstructions, thus fewer matches between \ac{ckm} and observations, which induces the plateaux-behavior in Fig.~\ref{fig:singleAuth}.

\begin{figure}
    \centering
    \input{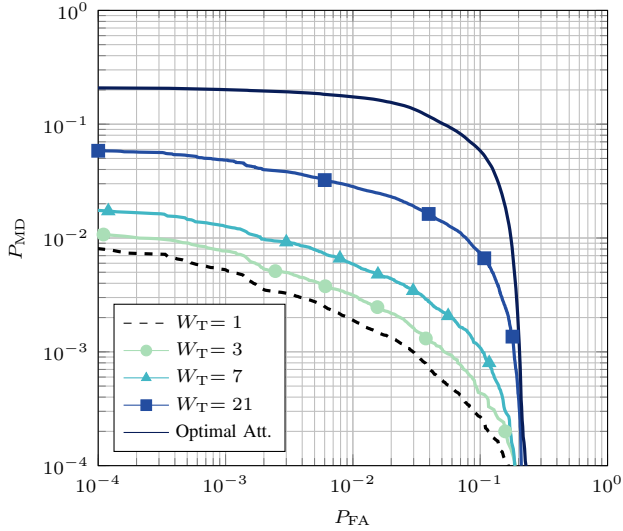}
    \caption{\ac{det} curve for $K = 0.05$ for considering random position attack, the informed attack with various values of $W_{\rm T}$, and the full map attack.}
    \label{fig:informedAttack}
    \vspace{-.5cm}
\end{figure}

Next, Fig. \ref{fig:informedAttack} shows the \ac{det} curves for the informed attack \eqref{eq:informedAtt}, including also the random position and the full map attack \eqref{eq:optimalAtt} against the \ac{lt}. From Trudy's point of view, the worst performance is obtained by using the random position attack, whereas the best is obtained by the full map attack.  As anticipated in Section~\ref{sec:attackStrategy}, these confirm to be lower and upper bounds of the informed attack, respectively.
Focusing on the informed attack, the performance improves as $W_\mathrm{T}$ increases. Still, when Trudy is moving as much as Alice, i.e.,  $W_\mathrm{T} = W_\mathrm{A}$, and knows Alice's previous position $\bm{p}_{k-1}$, only a limited improvement is shown, achieving $P_{\rm MD} \approx 2\cdot 10^{-2}$ instead of $P_{\rm MD} \approx 6\cdot 10^{-3}$ for $P_{\rm FA} = 10^{-2}$.
Notice that, signal obstructions lead to a constant value on the test function, for both legitimate and under attack scenarios, e.g., see \eqref{eq:detector}. Choosing a test threshold $\xi$ lower than this value immediately leads to $P_{\rm MD} = 0$ and $P_{\rm FA} = 1$, thus inducing a step behavior in the linear domain, which translates into the saturation-like behavior of the $P_{\rm FA}$ when in log scale, shown in Fig. \ref{fig:informedAttack}.

\begin{figure}
    \centering
    \input{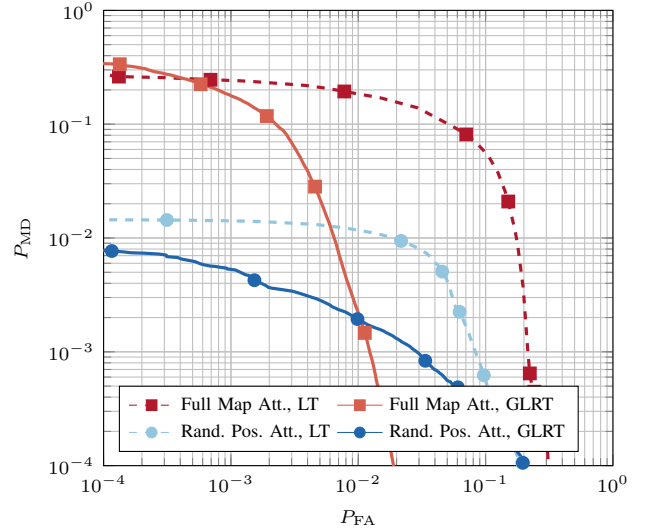}
    \caption{\Ac{det} comparing \ac{lt} and \ac{glrt} against random position and full map attack for $K = 0.05$.}
    \label{fig:multiBestAttack} 
    \vspace{-.5cm}
\end{figure}

In Fig.~\ref{fig:multiBestAttack}, we test random position and full map attacks against both \ac{lt} and \ac{glrt} checks. 
As expected, switching from the \ac{lt} to the \ac{glrt} makes the detector more vulnerable to the random position attack. 
In fact, consider a measurement obtained by the random attack that has a low likelihood $\gamma_{{\rm LR}}$ \eqref{eq:likelihood}, which would then be rejected by the \ac{lt}. This measurement may differ from the one selected by the full map attack, thus leading to a low denominator in \eqref{eq:GLRT}, and a higher overall test function, $\gamma_{\rm GLRT}$. Nevertheless, the performance of the full map attack decreases when using the \ac{glrt}. Still, even if the \ac{glrt} is specifically designed for detecting such an attack, this is not always accomplished. The \ac{glrt} in fact cancels measurements that are similar to the ones selected by the full map attack, thus interfering with Alice's measurements when $\bm{\phi}(\bm{p}^\star_{\rm T})$ is similar to $\hat{\bm{\phi}}^{(k)}$.

\section{Conclusions}\label{sec:conclusion}
We presented a novel \ac{pla} protocol that exploits the \acp{ckm} to authenticate a moving user in an indoor environment. 
Several attacks have been designed and include i) a random position attack, where the attacker randomly picks a position in the room and transmits from there, ii) an informed attack, where they exploit the \ac{ckm} to set their position choosing among the ones close to their current position, and iii) a full map attack, where the attacker can move instantly to any position of their choice.

Extensive numerical tests, with \acp{ckm} obtained via ray tracing, e.g., varying the number of available \acp{ap} or testing different estimation noise values, highlight the effectiveness of the proposed approach against all the considered attacks. In detail, for the case of a random-position attack, the system can achieve up to $P_{\rm MD}= P_{\rm FA} = 3\cdot 10^{-3}$. In the case of a full-map attack, the system has been shown to guarantee up to $P_{\rm MD}= P_{\rm FA} = 6\cdot 10^{-3}$.

\bibliographystyle{IEEEtran}
\bibliography{IEEEabrv,biblio.bib}

@STRING{IEEE_J_VT         = "{IEEE} Trans. Veh. Technol."}

@STRING{IEEE_J_ASSP       = "{IEEE} Trans. Acoust., Speech, Signal Process."}

@STRING{IEEE_J_COML       = "{IEEE} Commun. Lett."}

@STRING{IEEE_J_IFS        = "{IEEE} Trans. Inf. Forensics Security"}

@STRING{IEEE_J_CE         = "{IEEE} Trans. Consum. Electron."}

@STRING{IEEE_J_MI         = "{IEEE} Trans. Med. Imag."}

@STRING{IEEE_M_WC         = "{IEEE} Wireless Commun. Mag."}

@STRING{IEEE_O_CSTO        = "{IEEE} Commun. Surveys Tuts."}

@IEEEtranBSTCTL{IEEEexample:BSTcontrol,
CTLuse_forced_etal = "yes",
CTLmax_names_forced_etal = "6",
CTLnames_show_etal       = "1" }

@ARTICLE{Wang2024Knowledge,
  author={Wang, Qi and Liang, Wei and Zhang, Jialin and Wang, Ke and Jiang, Xiaolin},
  journal=IEEE_J_CE, 
  title={Knowledge-Enhanced Physical Layer Authentication for Mobile Devices}, 
  year={2024},
  month = {Nov.},
  volume={70},
  number={4},
  pages={7436-7448},
  doi={10.1109/TCE.2024.3379524}}

@ARTICLE{Baracca2012Physical,
  author={Baracca, Paolo and Laurenti, Nicola and Tomasin, Stefano},
  journal=IEEE_M_WC, 
  title={Physical Layer Authentication over {MIMO} Fading Wiretap Channels}, 
  year={2012},
  volume={11},
  month = {May},
  number={7},
  pages={2564-2573},
  doi={10.1109/TWC.2012.051512.111481}}

@article{Zhang2024Survey,
  author={Zhang, Junqing and Ardizzon, Francesco and Piana, Mattia and Shen, Guanxiong and Tomasin, Stefano},
  journal=IEEE_J_IFS, 
  title={Physical Layer-Based Device Fingerprinting For Wireless Security: From Theory To Practice}, 
  month = {May},
  year={2025},
  volume={20},
  number={},
  pages={5296-5325},
  doi={10.1109/TIFS.2025.3570118}}

@ARTICLE{Zeng2024Tutorial,
  author={Zeng, Yong and Chen, Junting and Xu, Jie and Wu, Di and Xu, Xiaoli and Jin, Shi and Gao, Xiqi and Gesbert, David and Cui, Shuguang and Zhang, Rui},
  journal=IEEE_O_CSTO, 
  title={A Tutorial on Environment-Aware Communications via Channel Knowledge Map for {6G}}, 
  year={2024},
  volume={26},
  number={3},
  month = {Feb.},
  pages={1478-1519},
  doi={10.1109/COMST.2024.3364508}}

@ARTICLE{Esrafilian2019Learning,
  author={Esrafilian, Omid and Gangula, Rajeev and Gesbert, David},
  journal=IEEE_J_IOT, 
  title={Learning to Communicate in {UAV}-Aided Wireless Networks: Map-Based Approaches}, 
  year={2019},
  volume={6},
  number={2},
  month = {Nov.},
  pages={1791-1802},
  doi={10.1109/JIOT.2018.2879682}}

@INPROCEEDINGS{Li2022Channel,
  author={Li, Kun and Li, Peiming and Zeng, Yong and Xu, Jie},
  booktitle={Proc. of Wireless Commun. and Netw. Conf. (WCNC)}, 
  title={Channel Knowledge Map for Environment-Aware Communications: {EM} Algorithm for Map Construction}, 
  year={2022},
  publisher = {IEEE},
  volume={},
  number={},
  pages={1659-1664},
  doi={10.1109/WCNC51071.2022.9771802}}

@ARTICLE{Haoyun2023Channel,
  author={Li, Haoyun and Li, Peiming and Cheng, Gaoyuan and Xu, Jie and Chen, Junting and Zeng, Yong},
  journal={J. Commun. Info. Netw.}, 
  title={Channel Knowledge Map ({CKM})-Assisted Multi-{UAV} Wireless Network: {CKM} Construction and {UAV} Placement}, 
  year={2023},
  volume={8},
  month = {Sept.},
  number={3},
  pages={256-270},
  doi={10.23919/JCIN.2023.10272353}}

@ARTICLE{Zhao2023Common,
  author={Zhao, Tianxiao and Wang, Xin and Yang, Wenfei and Xi, Xiaojun and Li, Jian},
  journal=IEEE_J_COML, 
  title={On the Common {AOA} Error in {CKM}-Based Integrated Sensing and Communications}, 
  year={2023},
  volume={27},
  number={7},
  pages={1859-1863},
  month = {Jul.},
  doi={10.1109/LCOMM.2023.3271356}}

@ARTICLE{Dai2024Prototyping,
  author={Dai, Zhuoyin and Wu, Di and Dong, Zhenjun and Li, Kun and Ding, Dingyang and Wang, Sihan and Zeng, Yong},
  journal=IEEE_J_VT, 
  title={Prototyping and Experimental Results for Environment-Aware Millimeter Wave Beam Alignment via Channel Knowledge map}, 
  year={2024},
  month ={Nov.},
  volume={73},
  number={11},
  pages={16805-16816},
  doi={10.1109/TVT.2024.3419795}}

@article{Hamza-2022, 
title={Developing three dimensional localization system using deep learning and pre-trained architectures for {IEEE} 802.11 {Wi-Fi}},
volume={4}, 
DOI={10.15587/1729-4061.2022.263185},
number={9(118)},
journal={East.-Eur. J. Enterp. Technol.}, 
author={Hamza, Aseel Hamoud and Hussein, Sabreen Ali and Ismaeel, Ghassan Ahmad and Abbas, Saad Qasim and Zahra, Musadak Maher Abdul and Sabry, Ahmad H.}, 
year={2022},
month={Aug.},
pages={41–47} }

@ARTICLE{ZhangChaoyue-2025,
  author={Zhang, Chaoyue and Zhou, Zhiwen and Xu, Xiaoli and Zeng, Yong and Zhang, Zaichen and Jin, Shi},
  journal=IEEE_J_VT, 
  title={Prototyping and Experimental Results for {ISAC}-Based Channel Knowledge Map}, 
  year={2025},
  volume={74},
  month = {Feb.},
  number={7},
  pages={10719-10731},
  doi={10.1109/TVT.2025.3545785}}

@article{Njeri-2014,
  title={Performance Analysis of {MUSIC}, Root-{MUSIC} and {ESPRIT} {DOA} Estimation Algorithm},
  author={Njeri P. Waweru and Dominic Bernard Onyango Konditi and Philip Kibet Langat},
  journal={Int. J. Electr. Comput. Energ. Electron. Commun. Eng.},
  year={2014},
  month={Mar.},
  volume={8},
  pages={209-216},
}

@book{Benvenuto-2021,
author = {Benvenuto, Nevio and Cherubini, Giovanni and Tomasin, Stefano},
year = {2021},
publisher = {John Wiley \& Sons},
title = {Algorithms for Communications Systems and their Applications},
isbn = {9781119567998},
doi = {10.1002/9781119567998}
}

@ARTICLE{Stoica-1989,
  author={Stoica, P. and Nehorai, Arye},
  journal=IEEE_J_ASSP, 
  title={{MUSIC}, maximum likelihood, and {Cramer-Rao} bound}, 
  year={1989},
  month={May},
  volume={37},
  number={5},
  pages={720-741},
  doi={10.1109/29.17564}}

@ARTICLE{Sijbers-1998,
  author={Sijbers, J. and den Dekker, A.J. and Scheunders, P. and Van Dyck, D.},
  journal=IEEE_J_MI, 
  title={Maximum-likelihood estimation of {Rician} distribution parameters}, 
  year={1998},
  month ={Jun.},
  volume={17},
  number={3},
  pages={357-361},
  doi={10.1109/42.712125}}

@ARTICLE{Ardizzon2025Relation,
  author={Ardizzon, Francesco and Tomasin, Stefano},
  journal=IEEE_J_COML, 
  title={On the Relation Between {OC-LSSVM} and Likelihood Test for Physical Layer Authentication}, 
  year={2025},
  volume={29},
  number={12},
  pages={2865-2869},
  month = {Oct.},
  doi={10.1109/LCOMM.2025.3617764}}
\end{document}